\documentstyle[aas2pp4]{article}

\newcommand {\hI} {\ion{H}{1}\,\,}
\newcommand {\kms} {\,km\,s$^{-1}$\,}
\newcommand {\M} {\mbox{${\cal M}$}}
\newcommand {\mhI} {\M$_{HI}$\,}
\newcommand {\msol} {\M$_\odot$\,}
\newcommand {\lsol} {\L$_\odot$\,}
\newcommand {\mhlb} {(\M$_{HI}$/L$_B$)\,}
\newcommand {\mhlv} {(\M$_{HI}$/L$_V$)\,}

\begin{document}
 
\title{Detection of \hI Associated with the Sculptor Dwarf Spheroidal Galaxy}
 
\author{Claude Carignan}
\affil{D\'epartement de physique and Observatoire du mont M\'egantic,
Universit\'e de Montr\'eal, C.P. 6128, Succ. centre ville,
Montr\'eal, Qu\'ebec, Canada. H3C 3J7\\
e--mail: carignan@astro.umontreal.ca}
 
\author{Sylvie Beaulieu}
\affil{Institute of Astronomy, University of Cambridge,
Madingley Road, Cambridge, England. CB3 0HA\\
e--mail: beaulieu@ast.cam.ac.uk}
 
\author{St\'ephanie C\^ot\'e}
\affil{Dominion Astrophysical Observatory,
Herzberg Institute of Astrophysics,
National Research Council of Canada,
5071 West Saanich Rd., Victoria, BC, Canada. V8X~4M6\\
e--mail: Stephanie.Cote@hia.nrc.ca}
 
\author{Serge Demers}
\affil{D\'epartement de physique and Observatoire du mont M\'egantic,
Universit\'e de Montr\'eal, C.P. 6128, Succ. centre ville,
Montr\'eal, Qu\'ebec, Canada. H3C 3J7\\
e--mail: demers@astro.umontreal.ca}
 
\author{Mario Mateo}
\affil{Departement of Astronomy, University of Michigan, 
821 Dennison Bldg., Ann Arbor,\\
 MI 48109--1090, USA.\\
e--mail: mateo@ra.astro.lsa.umich.edu}
 
\begin{abstract}
 
Neutral hydrogen
(\hI) has been detected in the Local Group dwarf spheroidal galaxy
Sculptor with the 64 m Parkes single dish and mapped with the Australia
Telescope synthesis array. Most of the detected \hI is in two clouds
$\sim 15'-20'$ away from the optical
center. The gas is observed at the same systemic velocity than the
stars but at $\ge 125$ \kms away from the Magellanic Stream components in that
region. A lower limit to the \hI mass of $3.0 \times 10^4$ \msol 
is derived from the
synthesis observation for an \mhlb $\simeq$ 0.02. This amount of \hI
is compatible with mass loss expected from normal giants even if
only 10\% of the gas is retained by the galaxy in its neutral form.
 
\end{abstract}
 
 
\section{INTRODUCTION}
 
It is generally believed that the dwarf spheroidal companions of the
Galaxy and of M31 are completely devoid of a detectable interstellar
medium (ISM), and in particular neutral hydrogen (eg Da Costa 1994).
Previous attempts to detect \hI\ in these galaxies (Knapp et al 1978;
Mould et al 1990; Koribalsky et al 1994) have been carried out with
the NRAO 91m, NRAO 41m and the Parkes 64m single dish telescopes;
these studies were only able to set upper limits of $\sim 10^2-10^5$
\msol\ for the \hI\ content of eight of the nine dSph companions of
the Milky Way.  Thuan and Martin (1979) set limits on the \hI\ content
of two of the lower luminosity dSph systems near M~31 (And~I and
And~III).
 
The only ambiguous detection near the Milky Way was the Sculptor
dSph system.  Knapp et al (1978; their figure 1) identified \hI\ while
observing Sculptor; they noted that
 
\begin{quotation}
 
``... it is therefore (remotely) possible that the galaxy is embedded
in the gas cloud at $+120$ \kms.'',
 
\end{quotation}
 
\noindent In the end, they dismissed this possibility because the
optical velocity of Sculptor was believed to be $\sim 200$ \kms\ at
that time.  Subsequent optical radial velocity studies of individual
stars in Sculptor have shown its systemic velocity to be $\sim 110$
\kms (Queloz et al 1995), quite close to
the \hI velocity reported by Knapp et al (1978).  Hence, in
retrospect is appears that Sculptor was the first dSph system in which
neutral gas was detected; however, because of the initial suggestion
that the \hI was not associated with the galaxy, this result has been
largely forgotten.
 
The ISM of dSph and dSph-like systems may be considerably more complex
than generally believed (see Mateo 1998 for a more extensive
discussion).  Johnson and Gottesman (1993) and Young and Lo (1997a)
found complex \hI\ distributions associated with the more
luminous dSph-like M~31 systems, NGC~185 and NGC~205.  In each case,
the gas is distributed asymmetrically about the optical galaxy.  21-cm
detections have also been reported in the remote Phoenix galaxy
(Carignan, Demers \& C\^ot\'e 1991), a system sometimes mentioned as a
transition system between dwarf irregular (dIrr) and dSph galaxies
(van de Rydt et al 1991; van den Bergh 1994).  Oosterloo et
al. (1996) found an \hI cloud near the even more remote Tucana dSph
galaxy, but concluded that the gas is more likely a high-velocity
cloud (HVC) associated with the Magellanic Stream (MS); this will
remain speculative until an optical radial velocity for this galaxy is
obtained.  
 
Sculptor is one of the closest of the Milky Way satellites at a
distance of 88 kpc (Kaluzny et al 1995).  Its angular size
(40\arcmin) is comparable to that of the apparently largest dSph
system (after Sagittarius), Fornax.  The distribution of stars along
the horizontal branch (HB) and the observed mean metallicity of
Sculptor imply that the bulk of the stars in the galaxy have ages
similar to that of the relatively young globular clusters found in the
outer halo (Kaluzny et al 1995).  The existing color-magnitude
diagrams (CMD) do not reveal a strong intermediate age population (Da
Costa 1984; Aaronson 1986), contrary to what is seen in Fornax or
Carina (Stetson 1997; Hurley-Keller et al 1998).  In contrast,
the substantial spread in metallicity seen in Sculptor (Da Costa 1988,
Kaluzny et al 1995) suggests that its star-formation (SF) history may
have been rather complex. Da Costa (1984) and Grebel et al (1994)
have noted a significant number of blue stars above Sculptor's main
sequence turn-off.  While Da Costa identified them as blue stragglers,
it is quite possible that these stars represent a small but
significant younger population.  Grebel et al (1994) argue strongly
that a second SF episode has occurred in Sculptor, possibly triggered
by the external accretion or internal accumulation of gas.  

The radial
velocity of Sculptor has been established by Queloz et al  (1995) to
be 109.9 $\pm 1.4$ \kms, confirming previous determinations of 109.2
$\pm 4.5$ \kms by Aaronson \& Olszewski (1987) and of 107.4 $\pm 2.0$
\kms by Armandroff \& Da Costa (1986).  The velocity dispersion,
derived from their sample of 23 K giants was found to be 6.2 $\pm 1.1$
\kms, for a global mass--to--light ratio $\sim 9 \pm 6$
(\M/L$_V$)$_\odot$.  The physical parameters of Sculptor are
summarized in Table 1.
%Table~\ref{tab1}.
 
There is ample, growing evidence that other galaxies in the Local
Group exhibit complex SF histories, but the relationship to their ISM
remains unclear (eg, see Mateo 1998).  As noted, Carina and Fornax
have complex SF histories, but no detected ISM.  Three more distant and
isolated objects, Phoenix (d $\leq 400$ kpc), Tucana (d $\leq 900$
kpc), and Antlia (d~$\simeq 1.15 \pm 0.1$~Mpc) share many of the
properties of the {\it bona fide} dSph systems. However, photometric
studies (Canterna and Flower, 1977; Ortolani and Gratton, 1988; van de
Rydt et al 1991; Whiting et al 1997) have
shown that Phoenix and Antlia possess stellar populations that seem to
share properties of the populations found in both dSph and dIrr
galaxies.  
 
The growing evidence of complex SF histories in virtually all dwarf
galaxies, including those with little or no detectable ISM, suggest
that we do not yet understand the relationship between the stellar and
gaseous components of these deceptively simple-looking systems (Mateo 1998).
Because Sculptor is so close and because of its past (though
unappreciated) detection at 21-cm, we have obtained new \hI
observations of the galaxy to study its ISM component in more detail.
The new radio observations are described in Section 2; these include
single dish \hI observations using the Parkes 64m telescope, synthesis
\hI mapping done with the Narrabri Australia Telescope Compact Array
(ATCA), and CO observations obtained with the Swedish-ESO
Submillimeter Telescope (SEST) in Chile.  In Section 3 we analyze in
detail the content and distribution of the \hI gas mapped with the
ATCA, while in Section 4 we speculate about the possible origin of the
detected gas in Sculptor.  We end with a summary of our principal
results and our main conclusions in Section 5.
 
\section{OBSERVATIONS}
 
\subsection{Single Dish Parkes Observations}
 
%(Table~\ref{tab2})
We observed Sculptor with the Parkes 64 m single dish telescope (Table
2) to confirm the detection reported by Knapp et al (1978).  These
new data were obtained in January 1992.  The flux calibrator PKS
1934--638 was used to measure a mean system temperature during the run
of T$_{sys}$ $\simeq 50$ K with a HPBW of 14\farcm 9 $\pm 0 \farcm
2$. The 8 MHz bandwith, centered at 300 \kms, was divided in 1024
channnels for a channel width of 1.65 \kms and a full velocity
coverage of [-544, 1144] \kms.
 
A total of 100 minutes integration was accumulated on Sculptor which,
after averaging the 12 individual 500 second observations, resulted in
a spectrum with an RMS noise of 0.013 Jy. This sets a 3$\sigma$
detection limit of $\sim$100 \msol\ at the distance of Sculptor.  As
we discuss in more detail below, we obtained a clear, strong detection
of Sculptor with these observations.  In addition to Sculptor, we also
obtained new observations of the relatively recently discovered
Sextans dSph (Irwin et al 1990).  Contrary to the Sculptor
observations, we did not detect \hI towards Sextans.  Our 3$\sigma$
upper limit for the neutral hydrogen content is 130 \msol for any gas
that might be located within the Parkes beam.  For both galaxies, the
Parkes telescope was centered directly on the optical centroids of the
galaxies during these observations.
 
\subsection{ATCA Synthesis Mapping}
 
%(Table~\ref{tab3}).
After confirming the earlier \hI detection by Knapp et al (1978), we
decided to try to map Sculptor with the 375m configuration of
the ATCA on 1992 October 2 (Table 3).  For these observations we
obtained a total of 8.2 hours of integration on source, in blocks of
35 minutes and separated by 5 minute integrations of PKS 0042--442 for
phase and amplitude calibration purposes. The absolute intensity scale
was set by a 30 minutes observation of PKS 0407--658, which we assumed
to have a flux density of 14.4 Jy at our observing frequency.
Normally, the full array is comprised of six antennae; however, data
from the 6th antenna located $\sim$6 km from the array center was
discarded during the reduction stage since no flux could be seen in
the longest baselines. This left 10 baselines with the remaining 5
antennas which spanned a baseline range of 61m to 459m. With this
configuration it is possible to detect structures up to
$\sim$22\arcmin. With the ATCA 22m antennae, the FWHM of the primary
beam is $\sim$33\arcmin. The mean system temperature during the run
was T$_{sys} \simeq 35$ K.  A bandwidth of 7.4 MHz was used with a
channel spacing of 7.8 kHz, for a velocity resolution of 1.65 \kms.
 
The synthesis data were edited and calibrated using the NRAO AIPS
reduction software package. A bandpass calibration was applied using
the PKS 0407--658 data and the continuum was subtracted by fitting a
straight line to the real and imaginary parts of line--free channels
in the UV plane and subtracting the fit from the data. The images were
then computed using natural weighting for the UV data (higher weight
to the shorter baselines) which gives maximum sensitivity while
decreasing the spatial resolution by about a factor of 2. During the
processing of generating the radio image, the data were also Hanning
smoothed in velocity, thus decreasing the velocity resolution to 3.3
\kms. The full resolution data cube has a synthesized beam of
$205\arcsec \times 86\arcsec$ and an rms noise in each channel of 2.5
mJy/beam. The CLEANed data cube, restored with a synthesized beam of
$240\arcsec \times 240\arcsec$, has a noise of 3.3 mJy/beam. The
channel maps of the cleaned data cube are shown in Figure 1.  One can
see already in these maps that most of the detected signal is located
away from the center of the image.  As with the Parkes observations,
the ACTA pointing corresponded to the optical centroid of Sculptor.
%Fig.~\ref{fig1}.
 
\subsection{SEST CO Observations}
 
%(Table~\ref{tab4})
Sculptor was observed with the SEST (Table 4) in the $J = 1
\rightarrow 0$ line emission of $^{12}$CO at 115 GHz in January
1997. The Schottky receiver and good weather conditions yielded system
temperatures averaging 370 K.  The low-resolution Acousto Optical
Spectrometer (LRS2) was used as the backend with a bandwidth of 1088
MHz and a frequency resolution of 1.4 MHz (3.6 \kms ), centered on a
heliocentric velocity of 110 \kms .  A detailed description of SEST
can be found in Booth et al (1989).
 
Antenna temperatures were calibrated using the chopper wheel method,
and relative calibration uncertainties are estimated to be about 10\%.
The beamwidth at this frequency is 43\arcsec.  The observations were
made in a double-beam-switching mode with a throw of 12\arcmin\ and a
scan integration time of 120 seconds. Integrations were performed at
five distinct locations on the galaxy, corresponding to the off-center
HI peaks in both the NE and SW components, as well as on the inner
edges of the HI clouds since this is very often where CO is detected
in dE systems (see p.e. NGC 185 \& NGC 205 in Young \& Lo, 1997a).
The spectra were co-added and smoothed in
various ways to try to detect any CO emission, but regardless of the
methods employed, no emission was found at any of these locations.
Spectra of all the observed positions were then added together,
yielding an average spectrum representing a total integration time of
15.5 hours, with a rms noise level of 2.1 mK.  The $3\sigma $~CO
brightness temperature limit is therefore $T_{mb} < 9.0$ mK, after
beam efficiency corrections (taking $\eta _{mb} = 0.7$ for the SEST at
115 GHz), corresponding to a CO integrated intensity $I_{CO} = \int
{T_{mb} dV} < 0.2$ K \kms .
 
\section{ANALYSIS OF THE HI DATA}
 
\subsection{HI Content}
 
%Fig.~\ref{fig2}
\ Figure 2 shows the average spectrum of the 100 minute Parkes
integration at the central position of Sculptor. From it, a systemic
velocity of 112 $\pm 3$ \kms was derived, very close to the optical
velocity of 110 $\pm 1.4$ \kms.  The detected flux of 5.7 Jy
translates into an \hI mass of $\sim 1.0 \times 10^4$ \msol at the
distance of Sculptor. Since, as will be seen below, a large fraction
of the \hI probably lies outside the HPBW of the Parkes beam, this mass
estimate should only be considered as a lower limit of the total \hI
content of Sculptor.
 
%Fig.~\ref{fig3}
\ Figure 3 shows the global profile from the ATCA observations.  Notice
that the systemic velocity from this spectrum is slightly different
from the Parkes spectrum: 102 $\pm 5$ \kms.  This already suggests
that a considerable fraction of the gas detected by the
synthesis observations is different from the gas seen in the the
single dish observation. This can be seen more clearly in Figure 4
where the two spectra have been superposed.  The integrated ATCA
profile gives a total detected flux of 17.3 Jy \kms, which implies a
total detected \hI mass of $3.0 \times 10^4$ \msol at the adopted
distance of Sculptor.
%Fig.~\ref{fig4}
 
\subsection{HI Distribution}
 
%Fig.~\ref{fig5}
\ Figure 5 shows the \hI surface density map for the gas detected by the
synthesis observation. It can be seen that most of the \hI is located
within two distinct clouds located $15'-20'$ from the optical center.
It is quite likely that a large fraction of the gas present in this
map was not detected in our single dish observations with the 14\farcm
9 Parkes beam centered at the optical position.  Moreover, a large
fraction of the gas is also outside the HPBW of the primary beam
($\sim 33'$) of the ATCA antennae; thus, even if the data in Figure 5
were corrected for primary beam attenuation, this single field
observation will have missed any gas located further out from the
center of the galaxy.  Given the observed distribution of gas in
Sculptor, it is at least plausible, and we believe probable, that
more extended \hI gas is present.
 
There is also another reason (already mentioned in Sec. 2.2) which
suggests that this observation may not have detected all the flux
present.  The diameter of the detected clouds is $\sim 20'$ which is
close to the largest structures ($\sim 22'$) that can be detected with
the shortest baseline ($\sim 60$m) of the 375m ACTA configuration we
employed.  Shorter baselines are needed to detect larger structures if
they exist.  Thus, the 3.0 $\times 10^4$ \msol of \hI detected within
the primary beam by the ATCA observation should also be considered as
a lower limit.  Another indication that the ATCA observations has not
detected all the flux can be seen in Figure 4 which shows that there
is some flux seen around 125 \kms in the single dish observations that
is not seen in the interferometric data.
%Fig.~\ref{fig4}
 
%Fig.~\ref{fig6}.
The fact that most of the \hI sits at the edge of the optical
component of Sculptor is illustrated even more clearly by the \hI
radial profile shown in Figure 6.  This distribution is reminescent of
what has been observed in some extreme dIrr systems. In their 1994 paper,
Puche and Westpfahl summarized the situation in low mass systems such
as Sextans A, Holmberg I and M81dwA as follows:
 
\begin{quotation}
 
``Very little gas remains in the central regions of the galaxies. The
inner limit of the \hI shell nearly coincides with the optical radius
of the galaxy. The outer extent of the \hI shell is about 1.5 times
the optical radius.''.
 
\end{quotation}
 
\noindent This is a nearly perfect description of what we have
observed in Sculptor, though a complete \hI shell has not yet been mapped.
 
\subsection{HI Kinematics}
 
%Fig.~\ref{fig7}.
The map of the isovelocity contours for the detected \hI is showed in
\ Figure 7.  The mean radial velocities for the 3 clouds are 98 $\pm 8$
\kms for the NE cloud, 119 $\pm 2$ \kms for the central cloud and 104
\kms $\pm 11$ for the SW cloud.  It is difficult to determine
conclusively whether the \hI clouds are in rotation around the main
body of the galaxy.  This might be surprising since no hint of
rotation has been found in any dSph systems that have adequate data
(Mateo 1994) with the exception maybe of Ursa Minor (Hargreaves et
al. 1994).  It is equally difficult to determine with any certainty
whether the clouds in Sculptor are systematically expanding or
contracting from or towards the galaxy center.  What makes the
interpretation of the kinematical data difficult is that we have no
independent constraint on the true orientation parameters of the
galaxy and even less on the origin or orientation of the gas. The
various different possibilities will be discussed in the next section.
 
%Fig.~\ref{fig8}.
The \hI velocity dispersion map is shown in Figure 8.  The mean
dispersions for the three clouds are 4.3 $\pm 2.5$ \kms for the NE
cloud, 2.3 $\pm 1.2$ \kms for the central cloud and 3.8 $\pm 2.2$ \kms
for the SW cloud with the exception of a small region (dark area
around 00$^{\rm h}$~59$^{\rm m}$~30$^{\rm s}$ and --33\arcdeg\
52\arcmin\ 00\arcsec) where $< \sigma > \simeq 15 \pm 2$ \kms.  
While the mean velocity dispersion
$\sim 4 \pm 2$ km s$^{-1}$ of the gas
in the outer parts appears to be smaller
than the stellar velocity dispersion $\sim 6 \pm 1$
km s$^{-1}$ observed in the center (Armandroff and Da Costa 1986;
Queloz et al 1995), they are still comparable within the quoted errors.
 
\section{DISCUSSION}
 
In trying to understand the properties of the ISM in Sculptor, we must
try to reconcile the following principal characteristics of the galaxy
and its relation to the Milky Way.  First, Sculptor is presently
located relatively close to the Milky Way, and its perigalactic
distance is even closer ($\sim 60$ kpc; Irwin and Hatzidimitriou 1995;
Schweizer et al 1995).  Consequently, it is possible (see sec. 4.2) that tidal
effects have played some role in producing the observed distribution
of gas in the galaxy.  This may help reconcile why gas that may be of
an internal origin, is located so close to the edge of the optical image
of the galaxy.
Because Sculptor is relatively close, one must also
carefully consider the possibility 
that the detected gas may not be associated with Sculptor but is instead
a high-velocity cloud from the Magellanic Stream or some other
complex that happens to be present in this part of the sky.
 
\subsection{Internal Origin}
 
Is it possible to account for the neutral gas seen in Sculptor from
mass loss in normal giants?  The most likely internal sources of gas
that can be realistically retained in the vicinity of Sculptor are
winds from evolved stars on the red--giant and asymptotic giant
branches, and gas expelled during the planetary nebula phase of
intermediate-age and old stars. Since the central regions of dSph
galaxies appear to be devoid of neutral gas (Knapp et al 1978; Mould
et al 1990; Koribalski et al 1994; this paper), and since they reveal
no obvious signs of dust or molecular gas (with the exceptions of
NGC~185 and NGC~205, both of which are sometimes considered
ultra-luminous dSph systems; Young and Lo 1997a; Mateo 1998), a
supernova would eject most, if not all, of the existing gas from a
galaxy as small as Sculptor (Mac-Low and Ferrara 1998) as well as its
own ejecta.   Any supernovae would simply complicate the gas-retention
problem; if a galaxy such as Sculptor is to retain gas from internal
sources, the more sedate (and slow) sources of gas must dominate the
generation of the ISM.
 
As summarized by Mould et al (1990), the total mass loss rate expected
from normal evolution is about 0.015 \msol yr$^{-1}$ per
$10^9$\lsol$_{,B}$.  For Sculptor ($L_B \sim 10^7$\lsol, Mateo 1998),
this implies a total return of $1.5 \times 10^5$\msol per Gyr.  At
this rate, it would take $\sim 200$ Myr to produce the observed amount
of \hI seen in Sculptor even if all of the ejected gas is retained by
the galaxy and is converted to neutral H, neither of which is probably
correct.  If we assume that the mass distribution in Sculptor is more
extended than the light (Da~Costa 1994), the escape velocity may be as
much as 3.0 times larger than the central dispersion of 6.6 km/s, or
about 20 km/s, or may be as only about twice the central dispersion,
or about 13 km/s, if the mass follows the light distribution.  The
velocity spectrum of red giant winds extends somewhat above even this
upper limit, suggesting that up to 80\%\ of the gas from such winds
can be lost from the galaxy.  Thus, the rejuvination time of the ISM
from internal sources must be considerably longer than 200 Myr.
 
\ For our purposes, the key point is that it should take from 200--1000
Myr to build up the amount of gas seen in Sculptor, and even longer if
a significant fraction of the gas is in molecular form
(observations of dE's suggest that there
could be as much mass in H$_2$ than in \hI; Wiklind, Combes \& Henkel 1995).
Since most of
the SF seems to have taken place between 8 and 10 Gyr in
Sculptor (Da Costa 1984), it would have produced a gas reservoir of
$\sim 3.0 \times 10^5$\msol. So, only 10\% of this need to be retained
in its neutral form to account for the \hI detected by the present
observations.
 
Of course, for other dSph galaxies in which \hI is not detected, these
same arguments should apply; for these systems the problem shifts to
how the gas is lost or is otherwise made unobservable.  Galaxies such
as Fornax (Stetson 1997; Demers et al 1998) and Carina (Smecker-Hane
et al 1994; Mighell 1997; Hurley-Keller et al 1998), which show clear
evidence of SF episodes even within the last few Gyr, show
as yet {\it no} evidence of neutral gas, at least in the central
regions.  In these cases, perhaps there simply has not been enough
time to generate a reservoir of gas.  But what about Ursa Minor, Draco,
Sextans, and Leo~II?   These galaxies contain no significant populations
younger than the youngest stars found in Sculptor.  If Sculptor could
retain gas from red giants and planetary nebulae, why didn't these?
Have we looked at the right place?
 
\subsection{Tidal Effects}
 
Another possibility is that the gas was removed from the outer parts
of Sculptor by tidal forces from the Milky Way during its last
perigalactic passage $\sim 10^8$ years ago (Irwin and Hatzidimitriou
1995).  This idea is based on the fact that while the central 10
arcmin of Sculptor (i.e. the optical core) has zero ellipticity,
outside this region, the ellipticity smoothly increased to the
asymptotic value of $\sim 0.3$ (Irwin and Hatzidimitriou 1995).
This picture is similar to numerical simulations of dSph galaxies that
are tidally disrupted where material is ejected ahead of and behind
the satellite (eg, Allen and Richstone 1988; 
McGlynn \& Borne 1991; Moore and Davis 1994;
Piatek and Pryor 1995; Oh et al 1995; Kroupa 1997).  Is it a
coincidence that the position angle of the proper motion measured by
Schweitzer et al (1995) of $40\arcdeg \pm 24\arcdeg$ happens to be
almost exactly the position angle defined by the two \hI clouds (see
\ Figure 5)?  
%Fig.~\ref{fig5}) ?
Tidal effects could produce two clouds
symmetrically distributed on both sides of the optical center.
Moreover, when stars (and gas) are detached from the host galaxy,
they continue to follow their host's galactic orbit for several
galactic years before dispersing beyond the host's tidal radius
(Oh  et al 1994).
 
With few exceptions, the dSph galaxies of the Local Group are
clustered around the MW and M31, while the dIrr galaxies are more
evenly distributed throughout the group (Mateo 1998).  This certainly
suggests that the proximity of a massive galaxy may have played an
important role in determining the structural and kinematic properties
of dSph systems.  The removal of their gas by tidal effects may be one
of the important results of these encounters, though whether this can
explain our observations of Sculptor remains unclear.
 
\subsection{Gas Falling Back or Expanding ?}
 
A number of authors have at various times suggested that dSph systems
may be the remnants of extremely low-luminosity dIrr galaxies that
have been depleted of gas by their initial or subsequent
SF episodes, or perhaps by tidal effects (eg, Ferguson and
Binggeli 1994; but also see Mateo 1998 and
references therein).  This scenario is supported by the fact that
large expanding cavities surrounded by dense shells are found in the
neutral interstellar medium  of many dIrr galaxies that were
observed with sufficient resolution (Puche and Westpfahl 1994).
The energetics of the gas suggest that these structures are
plausibly created by stellar winds and supernova explosions from the
young dIrr stellar populations (Larson 1974, Dekel \& Silk 1986).
 
The largest dwarfs, such as Magellanic irregular systems (e.g. IC
2574, Martimbeau, Carignan and Roy 1994; Holmberg II, Puche et
al. 1992), contain several such shells. However, in the smallest
dwarfs (e.g. Holmberg I and M81dwA, Westpfahl and Puche 1994; Leo A,
Young and Lo 1996), only one large slowly (v$_{exp} \simeq 5$ \kms)
expanding shell usually dominates the ISM.  The expansion and
contraction of the entire ring- or shell-like ISM of these small
galaxies is interpreted as being the result of burst(s)
of SF that took place in those systems.
 
dSph systems, such as Sculptor, are in the same mass range than the
low luminosity dIrr galaxies (M$_B \simeq -10$) and a similar process
may have taken place.  Soon after the primordial burst of SF, most of
the ISM could have been expelled from the inner regions of the galaxy
by stellar winds and supernova explosions, stopping SF.  Depending on
the energy released in the initial burst, and the total mass of the
system, some of the gas may be falling back into the galaxy, giving
the \hI a ring--like appearance. The ring size would depend on the
parent galaxy mass, the strength of the initial burst and time. In the
case of Sculptor, because of the missing short spacings of the present
observations which do not allow to see structures larger than $\sim
22'$, it is possible that we are only seeing the two regions of
highest surface density.  Perhaps most difficult for this interpretation
is the lack of any obvious, recent stellar population that may have 
ejected the gas.  As in the tidal scenario, we must suppose that the
gas, if ejected by SF processes, has been able to stay 
associated with Sculptor and remain in the outer parts of the galaxy
since its last burst.  
 
The \hI gas observed in a galaxy like Phoenix could have another
origin. Van de Rydt, Demers and Kunkel (1991) found that this system
has an intermediate population of blue stars with an age of $\sim 1.5
\times 10^8$ years along with an old, globular cluster-like population. If
that last burst of SF is responsible for the observed \hI location,
and if we assume a constant expansion velocity of 5 \kms (as estimated
in extreme low mass dIrr systems), most of it should be found around
$\sim 750$ pc ($\Delta \simeq 400$ kpc), which corresponds to $\sim$
6\farcm 5 on the sky.  
In recent VLA observations of Phoenix (Young \& Lo 1997b),
an \hI ``cloud'' was found at about the right distance west of the center.
However, it is difficult to know if that gas, detected at $\sim -23$
km s$^{-1}$, is truly associated with Phoenix since no optical velocity
is available.
 
If the same kind of calculations is applied to Sculptor, the radius
where the \hI is found would imply that the most recent burst of SF
would have taken place $\sim 10^8$ years ago. Could the blue stars
observed by Da Costa (1984) and Grebel et al (1994) around magnitudes
21 to 23 be the tip of such a population?  The deepest existing CMD of
Sculptor (Da Costa 1984) tells us that the most recent burst of SF
in the galaxy occurred more than 5-8 Gyr ago, but these data
are strictly from a small region near the center of Sculptor.  Thus,
while this scenario appears highly unlikely, deeper CMDs over a much
larger portion of Sculptor are needed to determine if a recent burst
of sufficient magnitude has occurred in this system. If such a
population were found, it could potentially explain the location and
amount of detected gas.  If no young stars (or too few) are present,
it becomes immediately more likely that if the 
observed \hI came from an internal source, the gas had
to have been produced during an initial 
burst of SF in Sculptor (then remained in the vicinity of the galaxy
for nearly a Hubble time), or else slowly accumulated from giants winds
and planetary nebulae as described above.
 
\subsection{External Origin}
 
A large \hI cloud was recently detected near the distant dSph galaxy
Tucana (Oosterloo et al 1996). Because the position of the cloud does
not coincide too well with the optical galaxy and because the total
gas mass inferred at Tucana's distance is rather large ($> 10^6$
\msol), these authors argued that the detected gas is more likely to
be a HVC associated with the MS (Mathewson et al 1974).  Could the
same thing be happening with Sculptor?  This is a reasonable
possibility: Sculptor is close to the South Galactic Pole where
the MS has a complex structure and several velocity components (Haynes
and Roberts 1979).  However, the highly symmetric distribution of the
gas in Sculptor relative to the optical image of the galaxy, and the
nearly perfect velocity correspondence between the radio and optical
observations, suggest strongly that in Sculptor's case we are not
dealing merely with an optical/radio illusion.  Instead, the proximity
of HVCs suggests that Sculptor may have actually accreted gas -- or is
in the process of doing so -- from external clouds.
 
There is circumstantial evidence supporting
the notion that dSph galaxies may be accreting gas of external origin.
In Carina, Smecker-Hane et al
(1994) argue for a large age spread yet a surprisingly small spread in
abundance.  This is in striking contrast to Leo~I which has a complex
SF history (Lee et al 1993) and a broad giant branch indicative of a
large abundance spread.  This behavior makes somewhat more sense if
the gas that formed distinct generations of stars in these galaxies
was accreted or captured from distinct clouds with their individual --
and therefore random -- mean abundances.  Wakker and van Woerden
(1997) review the observations of HVC in the halo; their summary of
past surveys suggest that many small, low--column density, and distant
clouds could still be hidden throughout the Galactic halo.
 
The low velocity dispersions that makes it difficult for dwarfs to
retain much of an ISM from internal sources also seems to limit the
feasibility of capture from outside sources even more severely.  The
dispersion in the halo is 10--20 times higher than in dwarfs such as
Sculptor, so very little gas could be captured in a random collision.
The other possibility is that the galaxies and clouds follow more
ordered motions.  Lynden-Bell and Lynden-Bell (1995) have recently
reviewed the possibility of kinematically and spatially related
streams of objects in the Galactic halo; they conclude that some such streams
may exist, but Sculptor in particular does not seem to be associated
with any stream containing any other galaxies or halo clusters.
Indeed, a measurement of the proper motion of Sculptor (Schweitzer et
al 1995) seems to rule out its association with all previously
identified galaxy streams.  Nonetheless, these results do not rule out
that Sculptor may be located in a stream of \hI clouds -- the lack of
distance information for the clouds makes it impossible to constrain
the existence of such streams.  If galaxies such as Sculptor (and
Carina, Fornax, and Leo~I when they were forming stars) did in fact
accrete gas from an external source, this would seem to be the only
way -- contrived as it is -- of doing so.
 
%However, based on the present data, we don't think that 
%the origin of the gas in Sculptor is external.
%First, the two clouds are exactly centered on the optical
%component and the probability that this happened for two foreground
%clouds is rather small. Second, the closest component of the MS, about
%one degree to the west, has a velocity V$_{GSR} \simeq -50$ \kms 
%(Haynes \& Roberts 1979) compared to V$_{GSR} \simeq 77$ \kms for Sculptor
%($\Delta V > 125$ \kms). It is thus unlikely, in this case,
%that the observed \hI gas could be HVC's associated with the MS
%but rather that it is truly associated to Sculptor.
 
\section{SUMMARY AND CONCLUSIONS}
 
We have obtained Parkes single dish and ATCA synthesis observations
of the Sculptor dSph galaxy and have clearly detected \hI associated
with this galaxy.  Our principal results include:
 
\noindent (1) The single dish observation has detected $1.0 \times
10^4$ \msol of \hI at a systemic velocity of $112 \pm 3$ \kms, which
is similar to the mean velocity of the stellar component.
 
\noindent (2) The synthesis observation allows to set a lower limit to
the total \hI content of $3.0 \times 10^4$ \msol. The mean velocity of
the gas detected with the ATCA is $102 \pm 5$ \kms.
 
\noindent (3) Most of the detected gas is located in two clouds
symmetrically distributed $15'-20'$ to the NE and SW of the optical
center. This distribution and good velocity coincidence with the
optical component of Sculptor virtually guarantees that the gas is
truly associated with the galaxy.
 
\noindent (4) The fact that a large fraction of the detected \hI is
outside the 33$'$ HPBW of the ATCA antennas and that the sizes of the
detected \hI clouds ($\sim 20'$) are close to the theoretical largest
structures ($\sim 22'$) that can be seen by the 375 m array
configuration, suggest that much more \hI may be present in Sculptor
but were missed by the present observations. The masses derived from
the interferometric observations should thus be considered as lower
limits.
 
\noindent (5) The amount of \hI detected is $\sim 10$\% of the
estimated mass loss from normal giants during its main epoch of SF
8-10 Gyr ago.
 
\noindent (6) Tidal effects due to the proximity of the MW may have
played a role in the observed distribution of the gas.
 
\noindent (7) The detected \hI could be either gas expelled in the
original burst of SF that is falling back onto the system or gas
expelled in a more recent burst of SF that would have taken place
$\sim 10^8$ years ago.  However, since there is as yet no clear
evidence for such a young stellar population in Sculptor, the first
alternative is only reasonable if there is a way for the gas to have
remained in the vicinity of Sculptor for most of the galaxy's
lifetime.
 
\noindent (8) If the gas is of an external origin, one must provide a
model where a low-mass galaxy such as Sculptor can acquire gas from
such a hot system as the Galactic halo.  Because the velocity
difference of Sculptor and the Magellanic Stream is $\sim 125$ \kms in
this part of the sky, it is very likely that the detected \hI did not
derive from that source of gas, but possibly from some other
unidentified or now-defunct stream.  In any case, the gas is almost
certainly not merely a component of the MS that is seen in projection
about the optical component of Sculptor.
 
%Table~\ref{tab5}
Table 5 compares the properties of the dSph systems to other Local
Group dIrr and dE systems. Looking at \mhlv it can be seen that
Sculptor and Phoenix appear to have about 10 times more \hI than has
been detected in the dE companions of M31 and about 40 to 150 times
less \hI than what is seen in dIrr galaxies having similar absolute
magnitudes. In the case of the dE galaxies, the \hI detected is
relatively close to the center. It is possible that more \hI is
present at larger radii. In the case of Sculptor and Phoenix, the \hI
masses are clearly lower limits.  It is possible that mapping larger
areas around those systems will reveal a much larger quantity of
gas. If this is the case, this would lend support to the suggestion
that some dSph systems may be remnants of extreme dIrr galaxies which
have been stripped of their gas by tidal effects or SF
bursts.  If this stripping/ejection scenario is correct, \hI should
also be observed at large radii in the other dSph systems.  Since the
observations used to derived the upper limits given in Table 5 were
centered on the optical images of the galaxies, it remains entirely
possible that other dSph systems contain large quantities of neutral
gas that has so far escaped detection.
%Table~\ref{tab5}
 
\acknowledgments
 
We would like to thank the staff of the Parkes and ATCA facilities for
their support during the HI data acquisition, James Lequeux for his
help in planning the SEST observations, and Tommy Wiklind for his
support during the CO observations. CC \& SD acknowledge grants from
NSERC.

\clearpage

\begin{deluxetable}{lr}
%\tablenum{1}
\tablewidth{25pc}
\tablecaption{Physical parameters of Sculptor.\label{tab1}}
\tablehead{}

\startdata
Morphological Type                              &dSph                   \nl
RA (J2000.0)\tablenotemark{a}  &01$^{\rm h}$ 00$^{\rm m}$ 09 \fs 4      \nl
Dec (J2000.0)\tablenotemark{a}          &--33\arcdeg 42\arcmin 33\arcsec\nl
l\tablenotemark{a}                              &287 \fdg 53            \nl
b\tablenotemark{a}                              &--83 \fdg 16           \nl
Galactocentric distance\tablenotemark{b}        &87.5 $\pm 6$ kpc       \nl
                                                &(1\arcmin\ $\simeq 25.5$ pc)\nl
Isophotal major diameter, D$_{25}$\tablenotemark{a}     &40\arcmin      \nl
Core radius, r$_c$\tablenotemark{c}             &5\farcm 8 $\pm 1 \farcm 6$\nl
Tidal radius, r$_t$\tablenotemark{c}            &76\farcm 5 $\pm 5 \farcm 0$\nl
Major axis PA\tablenotemark{c}                  &$99\arcdeg$ $\pm 1$\arcdeg\nl
Proper motion PA\tablenotemark{d}               &$40\arcdeg \pm 24\arcdeg$\nl
Abundances [Fe/H]\tablenotemark{b}              &[--2.2, --1.6]         \nl
Absolute magnitude, M$_B$\tablenotemark{a}      &--10.0                 \nl
Total luminosity, L$_B$                        &$1.5 \times 10^6$ L$_{\odot}$\nl
Absolute magnitude, M$_V$\tablenotemark{e}      &--10.7                 \nl
Total luminosity, L$_V$                        &$1.6 \times 10^6$ L$_{\odot}$\nl
Optical velocity, V$_{\odot}$\tablenotemark{e}  &$110 \pm 1.4$ \kms     \nl
Galactocentric velocity, V$_{GSR}$              &77 \kms                \nl
\tablenotetext{a}{de Vaucouleurs et al. (1991).}
\tablenotetext{b}{Kaluzny et al. (1995).}
\tablenotetext{c}{Irwin \& Hatzidimitriou (1995).}
\tablenotetext{d}{Schweitzer et al. (1995).}
\tablenotetext{e}{Queloz, Dubath \& Pasquini (1995).}
\enddata
\end{deluxetable}

\begin{deluxetable}{lr}
%\tablenum{2}
\tablewidth{25pc}
\tablecaption{Parameters of the Parkes HI observations.\label{tab2}}
\tablehead{}

\startdata
Dates of observations			&1992 January 25--30 		\nl
Flux calibrator				&PKS 1934--638			\nl
System temperature, T$_{sys}$		&$\sim$50 K				\nl
Primary beam at half--power (FWHM)	&14\farcm 9 $\pm 0 \farcm 2$	\nl
Bandwidth				&8.0 MHz			\nl
Central velocity			&300 \kms			\nl
Velocity range				&[-544, 1144] \kms		\nl
Channel width				&7.8 kHz			\nl 
					&(1.65 \kms)			\nl
\enddata
\end{deluxetable}

\begin{deluxetable}{lr}
%\tablenum{3}
\tablewidth{25pc}
\tablecaption{Parameters of the ATCA HI observations.\label{tab3}}
\tablehead{}

\startdata
Date of observations			&1992 October 2 		\nl
Integration time on source		&8.2 hours			\nl
Configuration				&375 m.				\nl
Baselines				&10 [61m, 459m]			\nl
Flux calibrator				&PKS 0407--658			\nl
Phase calibrator			&PKS 0042--442			\nl
System temperature, T$_{sys}$		&$\sim$35 K			\nl
Primary beam at half--power (FWHM)	&33\arcmin			\nl
Bandwidth				&7.4 MHz			\nl
Channel spacing (no smoothing)		&7.8 kHz			\nl
					&(1.65 \kms)			\nl
Channel spacing (after Hanning smoothing)&15.6 kHz			\nl 
					&(3.3 \kms)			\nl
FWHM of synthesized dirty beam          &$205\arcsec \times 86\arcsec$  \nl
FWHM of restored cleaned beam           &$240\arcsec \times 240\arcsec$ \nl
RMS noise in channel maps		&2.5 mJy/beam			\nl
(full resolution / no cleaning)		&\nl
RMS noise in channel maps		&3.3 mJy/beam			\nl
(after cleaning / convolved)		&\nl
Conversion factor, $240\arcsec \times 240\arcsec$ beam &0.01 K		\nl
(equivalent to 1 mJy/beam area)		&\nl
Maps gridding				&$30\arcsec \times 30\arcsec$ pixels\nl
\enddata
\end{deluxetable}

\begin{deluxetable}{lr}
\tablenum{4}
\tablewidth{25pc}
\tablecaption{Parameters of the SEST CO observations.\label{tab4}}
\tablehead{}

\startdata
Dates of observations			&1997 January 24--29		\nl
System temperature, T$_{sys}$		&370 K				\nl
Primary beam at half--power (FWHM)	&43\arcsec			\nl
Bandwidth				&1088 MHz			\nl
Central velocity			&110 \kms			\nl
Velocity range				&[-1300, 1500] \kms		\nl
Channel width				&1.4 MHz			\nl 
					&(3.6 \kms)			\nl
\enddata
\end{deluxetable}

\begin{deluxetable}{crrrrcc}
\tablenum{5}
\tablewidth{30pc}
\tablecaption{HI in Local dSph's, dI's \& dE's.\label{tab5}}
\tablehead{
\colhead{Name}
&\colhead{M$_V$\tablenotemark{a}}
&\colhead{R$_{GC}$\tablenotemark{a}}
&\colhead{L$_{tot}$}
&\colhead{\M$_{HI}$}
&\colhead{\mhlv}
&\colhead{ref.s}\\
\colhead{}
&\colhead{}
&\colhead{(kpc)}
&\colhead{$(L_V)_{\odot}$}
&\colhead{(\msol)}
&\colhead{(\M/L$_V$)$_{\odot}$}
&\colhead{HI data}
}

\startdata
\nl
 Fornax & --13.0 & 120 & $1.4 \times 10^7$ & $< 10^4$ & \nodata & (1) \nl
 Leo I & --11.5 & 198 & $3.4 \times 10^6$ & $< 10^4$ & \nodata &(1) \nl
 {\bf Sculptor} & --10.7 & 87.5\tablenotemark{b} & $1.6 \times 10^6$ &
 $\ge 3.0 \times 10^4$ & 0.02&(2)\nl
 {\it Phoenix}  & --9.9 & $\leq 400$ & $7.6 \times 10^5$ &
 $1.0 \times 10^5$ & 0.07 & (3) \nl
 Leo II & --9.6 & 207 & $5.9 \times 10^5$ & $< 10^4$ & \nodata & (1) \nl
 {\it Tucana}   & --9.5 & {\it $\leq 900$} & $5.3 \times 10^5$ &
 $1.5 \times 10^6$ & $\sim 3$ ? & (4) \nl
 Sextans & --9.2 &  83 & $4.1 \times 10^5$ & $< 130$ & \nodata & (5) \nl
 Carina& --8.6 &  85 & $2.4 \times 10^5$ & $< 10^3$ & \nodata & (6) \nl
 Ursa Minor & --8.4 &  64 & $2.0 \times 10^5$ & $< 280$ & \nodata & (1) \nl
 Draco& --8.3 &  72 & $1.8 \times 10^5$ & $< 68$ & \nodata & (1) \nl
 Sagittarius&\nodata&16\tablenotemark{c}&\nodata&\nodata&\nodata&\nodata\nl
 {\it Antlia}&\nodata&1150\tablenotemark{d}&\nodata&\nodata&\nodata&\nodata\nl
\nl
\tableline
\nl
 Leo A & --13.9 & 2200 & $5.3 \times 10^7$ & $8.1 \times 10^7$ & 1.5 & (7) \nl
 Sextans A & --13.8 & 1320 & $4.8 \times 10^7$ & $5.8 \times 10^7$ & 1.2 &(8)\nl
 M81dwA & --11.0 & 3250 & $3.6 \times 10^6$ & $1.1 \times 10^7$ & 3.0 & (9) \nl
 G.R. 8 & --10.6 & 1100 & $2.6 \times 10^6$ & $2.0 \times 10^7$ & 0.8 & (10) \nl
\nl
\tableline
\nl
 NGC 205& --15.7 &  850 & $1.6 \times 10^8$ & $4.3 \times 10^5$ & 0.003 &(11)\nl
 NGC 185& --13.8 &  600 & $2.8 \times 10^7$ & $1.0 \times 10^5$ & 0.004 &(11)\nl
\nl
\tablenotetext{a}{Irwin \& Hatzidimitriou (1995).}
\tablenotetext{b}{Kaluzny et al. (1995).}
\tablenotetext{c}{Ibata, Gilmore \& Irwin (1995).}
\tablenotetext{d}{Whitting, Irwin \& Hau (1997).}
\tablerefs{(1) Knapp, Kerr \& Bowers 1978;
(2) this paper (ATCA);
(3) Carignan, Demers \& C\^ot\'e 1991;
(4) Oosterloo, Da Costa \& Staveley--Smith 1996;
(5) this paper (Parkes);
(6) Mould et al. 1990;
(7) Young \& Lo 1996;
(8) Skillman et al. 1988;
(9) Sargent, Sancisi \& Lo 1983;
(10) Carignan, Beaulieu \& Freeman 1990;
(11) Young \& Lo 1997a}
\enddata
\end{deluxetable}

\clearpage
 
\begin{figure}
\figurenum{1}
\plotone{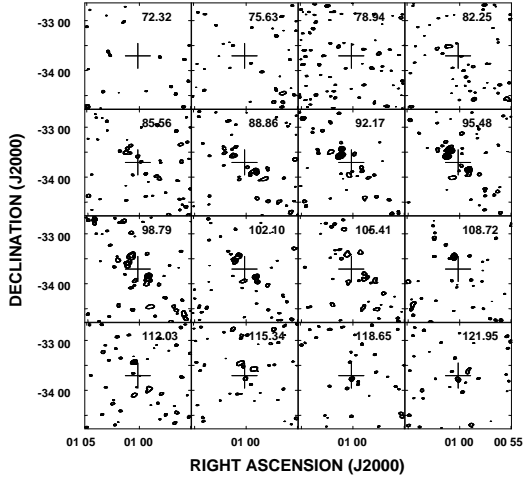}
\caption
{Channels maps from the AT observations. The levels are -0.1, 0.1, 0.2,
0.3, 0.4 \& 0.5 K. The synthesized beam is shown in the bottom left corner
of the first channel. The  velocity of each channel is given at the top right
corner. The cross in the center is 30\arcmin $\times$ 30\arcmin.}
\end{figure}
 
\begin{figure}
\figurenum{2}
\plotone{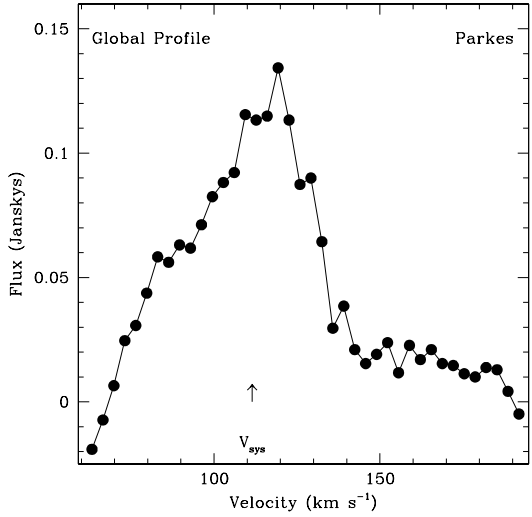}
\caption
{Global profile from the single dish Parkes observations. From this
profile, one gets V$_{sys} = 112$ \kms and \mhI $\simeq 1 \times 10^4$ \msol.}
\end{figure}

\begin{figure}
\figurenum{3}
\plotone{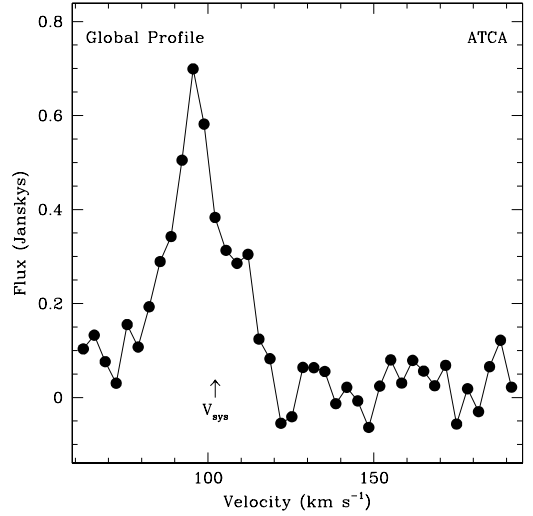}
\caption
{Global profile from the AT synthesis observations. From this
profile, one gets V$_{sys} = 102$ \kms and \mhI $\simeq 3 \times 10^4$ \msol.}
\end{figure}
 
\begin{figure}
\figurenum{4}
\plotone{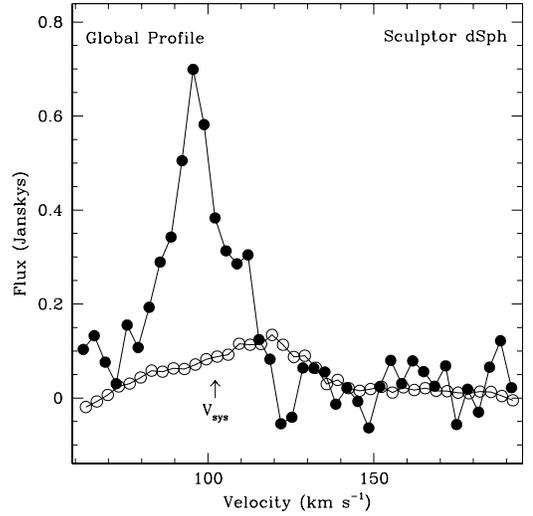}
\caption
{Global profile derived from the Parkes observations superposed on the
global profile obtained from the AT observations.}
\end{figure}
 
\begin{figure}
\figurenum{5}
\plotone{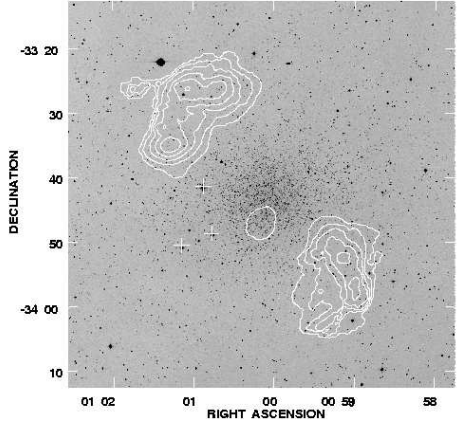}
\caption
{HI surface densities superposed on the STScI
Digitized Sky Survey optical image. The countours
are 0.2, 0.6, 1.0, 1.4, 1.8 \& 2.2 $\times 10^{19}$ cm$^{-2}$. The data
have been corrected for the primary beam attenuation. }
\end{figure}
 
\begin{figure}
\figurenum{6}
\plotone{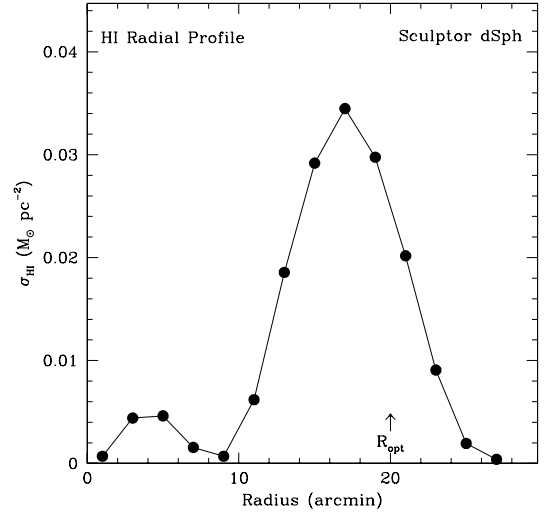}
\caption
{HI surface density radial profile. The optical size (R$_{opt}$) is indicated.}
\end{figure}
 
\begin{figure}
\figurenum{7}
\plotone{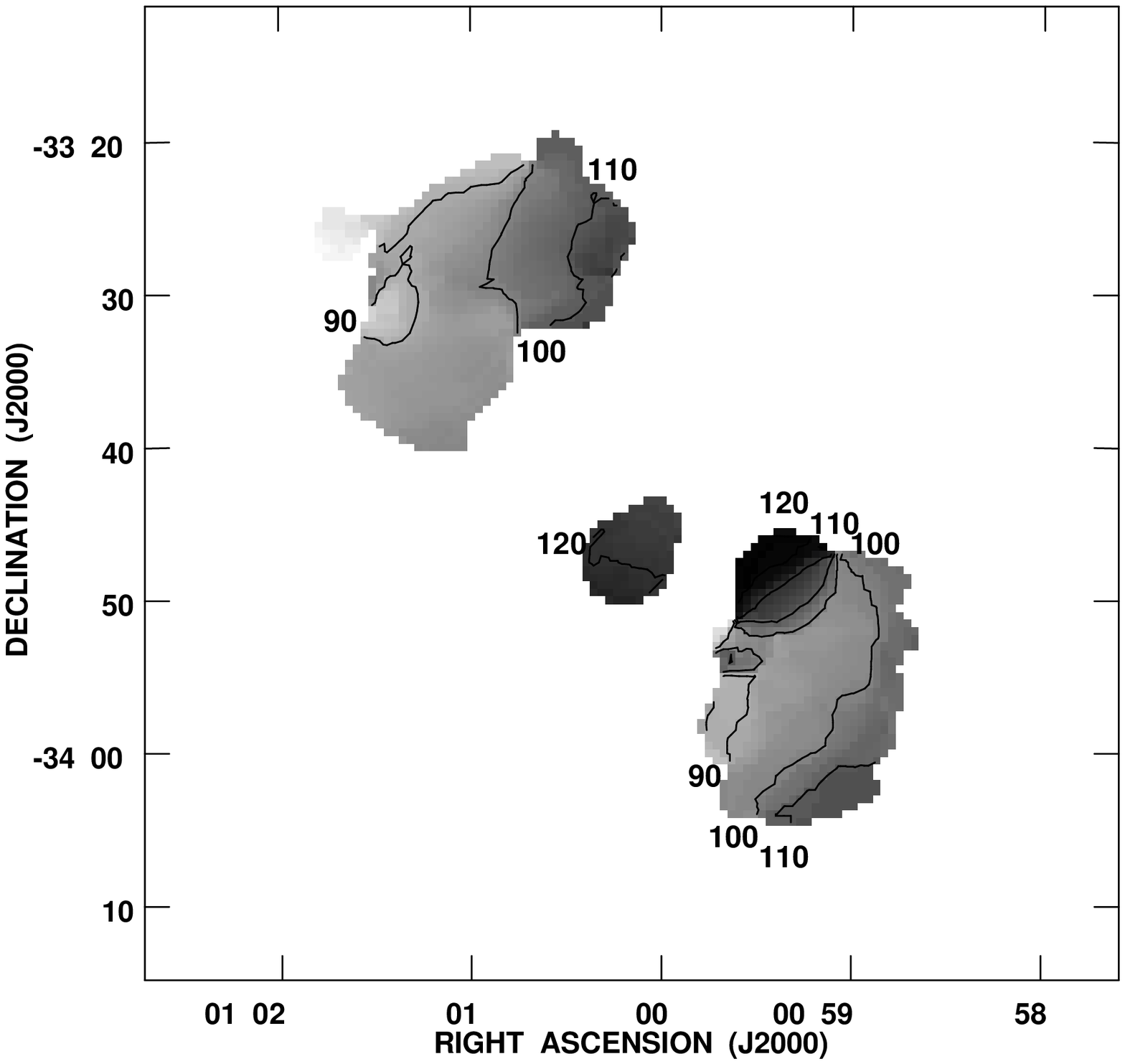}
\caption
{Velocity field map. The countours are 90, 100, 110, 120 \kms.}
\end{figure}
 
\begin{figure}
\figurenum{8}
\plotone{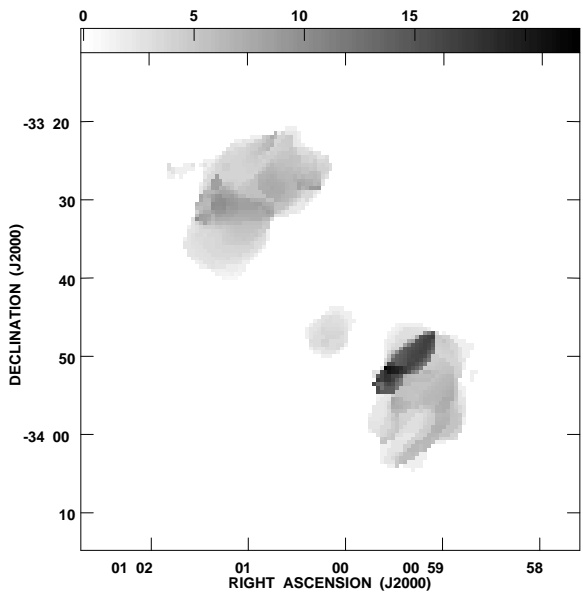}
\caption
{Velocity dispersion map. The velocity dispersion of the gas is almost
everywhere less than 10 \kms. The mean dispersion is $<\sigma> \simeq 2.3
\pm 1.2$ \kms in the central cloud and $<\sigma> \simeq 4.1 \pm 2.3$ \kms
in the NE \& SW clouds with the exception of a region of high $\sigma$
($<\sigma> \simeq 15 \pm 2$ \kms) in the SW cloud.}
\end{figure}
 
\end{document}